\documentclass{cernrep}
\begin{document}
\title{QCD under extreme conditions: an informal discussion}
\author{E.S. Fraga\thanks
                 {On leave from Instituto de F\'isica, Universidade Federal do Rio de Janeiro,
Rio de Janeiro, RJ, Brazil.}}
\institute{Institute for Theoretical Physics, J.~W.~Goethe-University, Frankfurt am Main, Germany\\
Frankfurt Institute for Advanced Studies, J. W. Goethe University, Frankfurt am Main, Germany}
\maketitle

\begin{abstract}
We present an informal discussion of some aspects of strong interactions under extreme conditions 
of temperature and density at an elementary level. This summarizes lectures delivered at the 2013 CERN -- Latin-American 
School of High-Energy Physics and is aimed at students working in experimental high-energy physics.
\end{abstract}

\section{Introduction and motivation: why, where and how}

Quantum Chromodynamics (QCD) is an extremely successful theory of strong interactions that has passed numerous tests 
in particle accelerators over more than 40 years \cite{QCD-general}. This corresponds to the behavior of hadrons in the vacuum, 
including not only the spectrum but also all sorts of dynamical processes. More recently strong interactions, and therefore QCD, has 
also started being probed in a medium, under conditions that become more and more extreme \cite{QCD-extreme}. Although quite 
involved theoretically, this is not just an academic problem. In order to make it clear, one should consider three very basic questions, 
that should always be asked in the beginning: why? where? how?

\subsection{Why?}

It was realized since the very beginning that strong interactions exhibit two remarkable features that are related but represent 
properties of complementary sectors of the energy scale. The first one is asymptotic freedom \cite{asymptotic-freedom}, which 
can be perturbatively demonstrated by an explicit computation of the beta function to a give loop order in QCD \cite{QCD-beta-function}. 
The second, which is consistent with the first but should be seen as totally independent, since it is a property of the nonperturbative 
vacuum of strong interactions, is color confinement \cite{confinement}. Even though reality constantly shows that confinement is a 
property of strong interactions, and therefore should somehow be built in QCD, this proof remains a theoretical open problem so far. 
Even for the pure Yang-Mills theory, where the bound states correspond to glueballs, the existence of a mass gap is still to be shown after 
more than half a century of the original paper on nonabelian gauge theories \cite{YM}. For this reason, confinement is ranked in the 
Clay Mathematics Institute list of unsolved Millennium problems \cite{clay}.

Much more than a cute (and very tough) mathematical problem, this is certainly among the most important theoretical and 
phenomenological problems in particle physics, since hidden there is the real origin of mass, as we feel in our everyday lives and 
experience with ordinary (and not so ordinary) matter. Although the Higgs mechanism provides a way to give mass to elementary 
particles in the Standard Model \cite{SM}, most of what constitutes the masses of hadrons come from interactions. 
For instance, more than $90\%$ of the proton mass originates in quark and gluon condensates \cite{condensates}. So, in spite 
of the fantastic success of the Standard Model \cite{SM}, we do not understand a few essential mechanisms.

Extremely high temperatures and densities bring us to an energy scale that facilitates deconfinement, and matter under such 
extreme conditions can behave in unexpected ways due to collective effects. This is, of course, a way to study the mechanism 
of confinement (by perturbing or modifying this state of matter). This leads us also to a deeper yet childish motivation, that of 
understanding what happens if we keep making things hotter and hotter, or keep squeezing things harder and harder \cite{wilczek-PT}. 
These questions can be reformulated in a more technical fashion as 'what is the inner structure of matter and the nature of strong 
interactions under extreme conditions of temperature and density?'. In experiments, one needs to ``squeeze'', ``heat'' and ``break''. 
From the theoretical point of view, one needs a good formulation of in-medium quantum field theory, using QCD or effective theories.

It is clear that the challenge is enormous. Although confinement seems to be a key feature of hadrons, and manifests also in relevant scales 
such as $f_{\pi}$ or $\Lambda_{QCD}$, it only {\it seems} to be present in QCD. So far, controlled lattice simulations show strong 
evidence of confinement in the pure gauge theory \cite{lattice-YM-conf}. As hinted previously, however, the theory is nonperturbative at the relevant scales, so that analytic methods are very constrained. And, although lattice simulations have developed to provide solid results
in several scenarios, they are not perfect. And, more important, they are not Nature. To make progress in understanding, or at least collecting 
important facts, one needs it all: experiments and observations, lattice simulations, the full theory in specific (solvable to some extent) limits 
and effective models. And also combinations, whenever possible, to diminish the drawbacks of each approach. 

\begin{figure}[ht]
\begin{center}
\includegraphics[width=8cm]{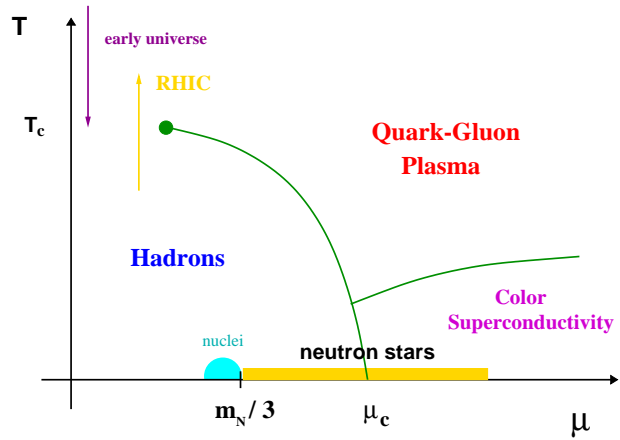}
\caption{Cartoon of a phase diagram for strong interactions. Extracted from Ref. \cite{Fraga:2003uh}}
\label{fig:cartoon}
\end{center}
\end{figure}

Whichever the framework chosen, collective phenomena will play a major role. Although somewhat put aside in the so-called 
microscopic ``fundamental'' particle physics, collective effects can affect dramatically the behavior of elementary particles in a medium 
under certain conditions. Besides the well-known examples of BCS and BEC phases in condensed matter systems \cite{BCS-BEC}, 
and also in dense quark matter \cite{colorSUC}, it was recently found that photons can form a Bose-Einstein condensate \cite{photons-BEC}. 
In fact, the textbook case of water and its different phases is quite illustrative of the richness that comes from collective phenomena 
that would hardly be guessed from the case of very few or non-interacting elementary particles.

In terms of the thermodynamics, or many-body problem, the basic idea is to perturb the (confined) vacuum to study confinement by 
heating (temperature), squeezing or unbalancing species (chemical potentials for baryon number, isospin, strangeness, etc) and 
using classical external fields (magnetic, electric, etc), so that the system is taken away from the confined phase and back. One can also 
relate (or not) confinement to other key properties of strong interactions, such as chiral symmetry. And, from the theorist standpoint, draw 
all possible phase diagrams of QCD and its ``cousin theories'' (realizations of QCD with parameters, such as the number of colors or flavors, 
or the values of masses, that are not realized in Nature) to learn basic facts. There are several examples, one well-known being the 
`Columbia plot', where one studies the nature of the phase transitions and critical lines on the $(m_{u}=m_{d},m_{s})$ plane. Nevertheless, 
if one draws a cartoon of the phase diagram in the temperature vs. quark chemical potential, for instance Fig. \ref{fig:cartoon}, 
and compares it to computations from effective models, lattice simulations and freeze-out points extracted from high-energy heavy ion 
collision data, one sees that the points still scatter in a large area \cite{stephanov-review}. So, there is still a long way ahead.

\subsection{Where?}

According to the Big Bang picture and the current description of the evolution of the early universe \cite{early-universe}, 
we expect that at about $10^{-5}s$ after the Big Bang a soup of quark-gluon plasma (in the presence of electrons, photons, etc) 
has undergone a phase transition to confined hadrons. This was, of course, the first realization of a QCD transition. This process 
was thermally driven and happened at very low baryon chemical potential.

It is quite remarkable that the scales of strong interactions allow for the experimental reproduction of analogous conditions in 
high-energy ultra-relativistic heavy ion collisions in the laboratory \cite{HIC}. In a picture by T. D. Lee, these collisions are seen 
as heavy bulls that collide and generate new states of matter \cite{bulls}. Such experiments are under way at 
BNL-RHIC \cite{BNL-RHIC} and CERN-LHC \cite{CERN-LHC}, and will be part of the future heavy ion programs at 
FAIR-GSI \cite{FAIR-GSI} and NICA \cite{NICA}.

For obvious reasons, it is common to refer to such experiments as ``Little Bangs''. However, one should be cautious with this point. 
In spite of the fact that the typical energy scales involved need to be the same, as well as the state of matter created, the so-called 
quark-gluon plasma \cite{QGP}, the relevant space-time scales differ by several orders of magnitude. Using a simple approximation 
for the equation of state, 
\begin{equation}
3p\approx \epsilon \approx \frac{\pi^{2}}{30}N(T) T^{4} ~,
\end{equation}
where $p$ is the pressure, $\epsilon$ the energy density and $N(T)$ the number of relevant degrees of freedom, 
we can easily estimate the typical sizes involved. The radius of the universe at the QCD phase transition epoch, as given 
by the particle horizon in a Robertson-Walker space-time \cite{weinberg}, where the scale factor grows as $a(t)\sim t^{n}$, 
is given by ($n=1/2$ and $N(T)\sim 50$ at this time for QCD)
\begin{equation}
L_{\rm univ}(T)\approx \frac{1}{4\pi} \left( \frac{1}{1-n} \right) \left(\frac{45}{\pi N(T)}\right)^{1/2} \frac{M_{\rm Pl}}{T^{2}} =
\frac{1.45 \times 10^{18}}{(T/{\rm GeV})^{2}\sqrt{N(T)}} {\rm fm} ~.
\end{equation}
Here $M_{\rm Pl}$ is the Planck mass, and it is clear that the system is essentially in the thermodynamic limit. 

\begin{figure}[ht]
\begin{center}
\includegraphics[width=12cm]{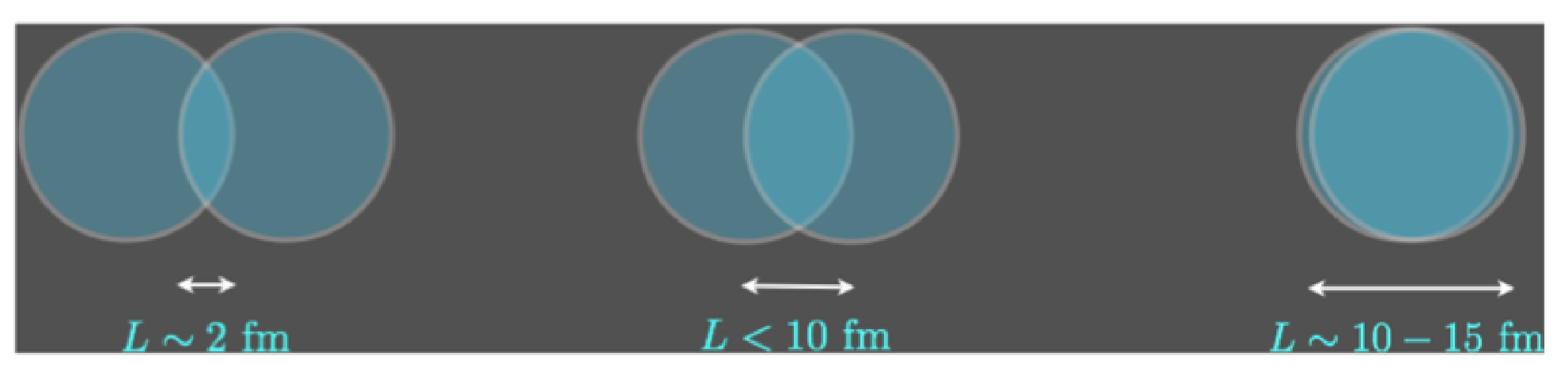}
\caption{Cartoon representing non-central heavy ion collisions and how they affect the size of the system.}
\label{fig:noncentral}
\end{center}
\end{figure}

On the other hand, in heavy ion collisions the typical length scale of the system is $L_{\rm QGP} \lesssim 10-15~$fm, 
so that the system can be very small, especially if one considers non-central collisions \cite{finite} 
(see Fig. \ref{fig:noncentral}). 
One can develop analogous arguments for the time scales given by the expansion rates, finding that the whole process in 
the early universe happens adiabatically, whereas in heavy ions it is not even clear whether the system can achieve thermal 
equilibrium, given the explosive nature of the evolution in this case. So, there are certainly large differences (in time and length scales) 
between Big and Little Bangs...

Keeping this caveat in mind, heavy ion experiments have been investigating new phases of matter at very high energies for 
more than a decade, producing an awesome amount of interesting data and a richer picture of strong interactions (see 
Ref. \cite{review-HIC} for a review).

In the realization of the Big and Little Bangs one is always in the high temperature and low density (small baryon chemical 
potential) sector of the phase diagram of strong interactions. However, high densities (at very low temperatures) can also 
probe new states of hadronic matter, and that is what is expected to be found in the core of compact stars \cite{compact-stars}. 
There, new phases, condensates and even color superconductivity may be present. In particular, the deconfinement and chiral 
transitions might affect significantly the explosion mechanism in supernovae \cite{compact-stars} via modifications in the equation of state. 

After a neutron (or hybrid) star is formed, densities in its core can in principle reach several times the nuclear saturation 
density $n_{0}=0.16~ {\rm fm}^{-3}=3 \times 10^{14} {\rm g/cm}^{3}$, which corresponds to squeezing $\sim 2$ solar masses into 
a sphere of $\sim 10$ km of radius. To describe these objects, one needs General Relativity besides in-medium quantum field theory.

\subsection{How?}

The reader is hopefully already convinced that, in order to describe the phenomenology of the phase structure and dynamics 
of strong interactions under extreme conditions, one needs all possibilities at disposal: theory, effective modeling, etc. We do not 
have one problem ahead, but a myriad of different problems. So, one has to make a choice. Our focus here will be the equation of 
state, of which we will discuss a few aspects. 

At this point, we are lead again to the ``why'' question. And the answer is because, besides carrying all the thermodynamic 
equilibrium information we may be interested in, it is also the basic crucial ingredient for dynamics, structure, etc. In fact, the phase diagram 
topology is determined in every detail by the full knowledge of the pressure $p(T,\mu,B,\dots)$. This will determine all phases 
present as we dial different knobs, or control parameters, such as temperature or chemical potentials.

The structure of a compact star, for instance, is given by the solution of the Tolman-Oppenheimer-Volkov (TOV) 
equations \cite{compact-stars}, which encode Einstein's General Relativity field equations in hydrostatic equilibrium 
for a spherical geometry:
\begin{eqnarray}
\frac{dp}{dr}&=&-\frac{G{\cal M}(r)\epsilon(r)}{r^2 \left[ 1-\frac{2G{\cal M}(r)}{r}  \right]}
\left[ 1+\frac{p(r)}{\epsilon(r)}  \right]
\left[ 1+\frac{4\pi r^3 p(r)}{{\cal M}(r)}  \right] ~,\\
\frac{d{\cal M}}{dr}&=&4\pi r^2 \epsilon(r)   \;\; ;\;\;\; {\cal M}(R)=M ~.
\end{eqnarray}
Given the equation of state $p=p(\epsilon)$, one can integrate the TOV equations from the origin until the pressure
vanishes, $p(R)=0$. Different equations of state define different types of stars (white dwarfs, neutron stars, strange stars, 
quark stars, etc) and curves on the mass-radius diagram for the families of stars.

\begin{figure}[ht]
\begin{center}
\includegraphics[angle=90,width=10cm]{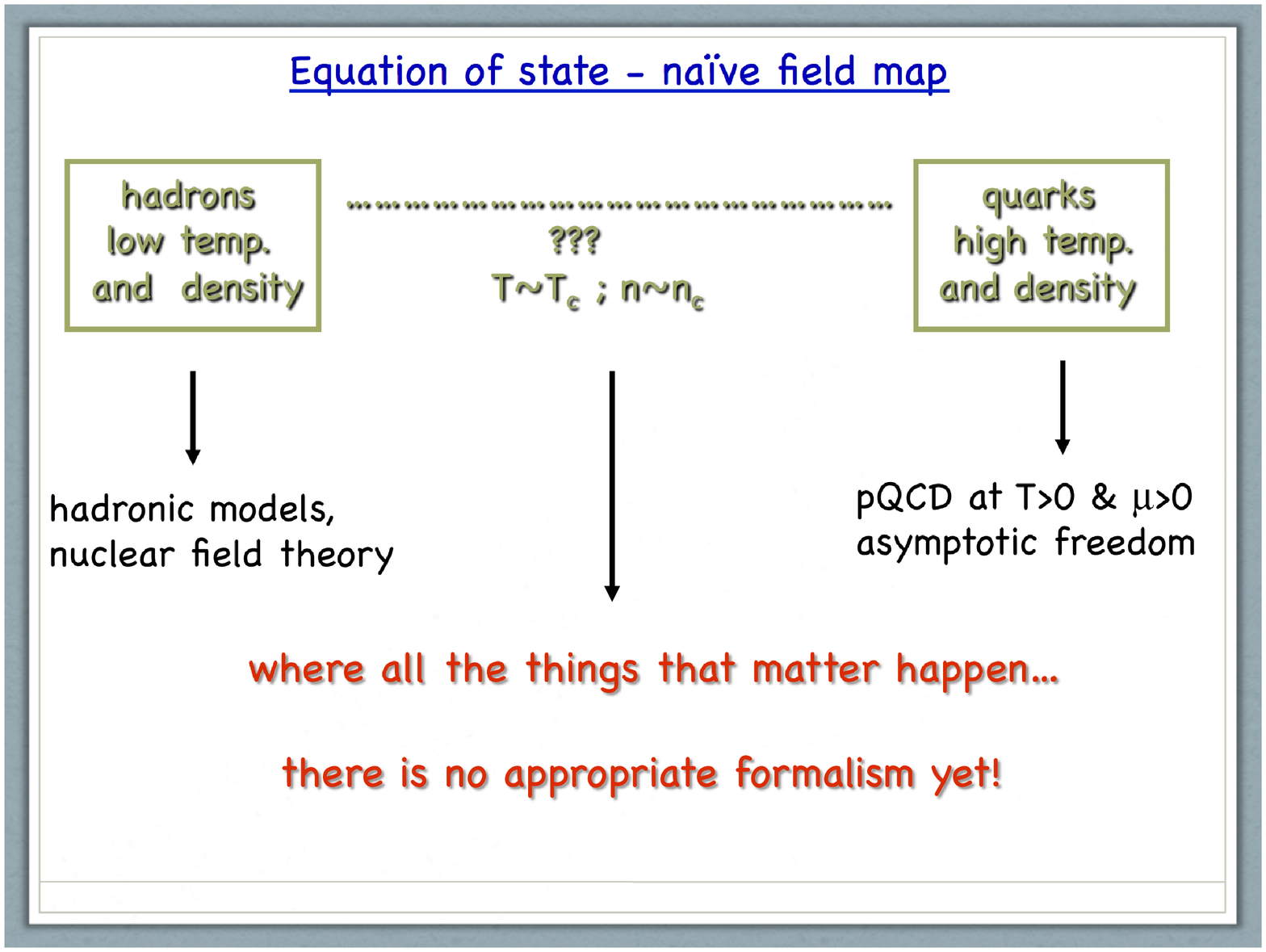}
\caption{Cartoon of the na\"\i ve field map for the equation of state for strong interactions.}
\label{fig:field-map}
\end{center}
\end{figure}

Furthermore, to describe the evolution of the hot plasma created in high-energy heavy ion collisions, 
one need to make use of hydrodynamics, 
whose fundamental equations encode the conservation of energy-momentum ($\partial_{\mu}T^{\mu\nu}=0$) and of baryon 
number (or different charges) ($\partial_{\mu}n_{B}v^{\mu}=0$, with $v^{\mu}v_{\mu}=1$). These represent only five equations for 
six unknown functions, the additional constraint provided by the equation of state. Hence, it is clear that we really need the 
equation of state to make any progress.

In principle, we have all the building blocks to compute the equation of state. The Lagrangian of QCD is given, so one would 
have ``simply'' to compute the thermodynamic potential, from which one can extract all relevant thermodynamic functions. The fact 
that the vacuum of QCD is highly nonperturbative, as discussed previously, makes it way more complicated from the outset. 
As we know, QCD matter becomes simpler at very high temperatures and densities, $T$ and $\mu$ playing the role of the momentum 
scale in a plasma, but very complicated in the opposite limit. On top of that, $T$ and $\mu$ are, unfortunately, not high enough in the 
interesting cases, so that the physically relevant region is way before asymptotic freedom really kicks in. Perturbative calculations 
are still an option, but then one has to recall that finite-temperature perturbative QCD is very sick in the infrared, and its na\"\i ve 
formulation breaks down at a scale given by $g^{2}T$ \cite{FTFT}. This is known as Linde's problem: at this scale, for a $(\ell+1)$-loop 
diagram for the pressure, for $\ell>3$ all loops contribute to the term of order $g^{6}$ even for weak coupling \cite{FTFT}.

The situation does not look very promising, as illustrated by the cartoon of Fig. \ref{fig:field-map} which shows that there is no appropriate 
formalism to tackle with the problem in the physically relevant region for the phase structure, namely the critical regions. However, 
there are several ways out. Some popular examples being: very intelligent and sophisticated ``brute force'' (lattice QCD), 
intensive use of symmetries (effective field theory models), redefining degrees of freedom (quasiparticle models), 
``moving down'' from very high-energy perturbative QCD, ``moving up'' from hadronic low-energy (nuclear) models. And we can and 
should also combine these possibilities, as discussed previously.

\section{Symmetries of QCD and effective model building}

\subsection{The simplest approach: the bag model}

Before discussing the building of effective models based on the symmetries, or rather approximate symmetries, of QCD, let us 
consider a very simple description: the MIT bag model \cite{FTFT} applied to describe the thermodynamics of strong interactions.

The model incorporates two basic ingredients, asymptotic freedom and confinement, in the simplest and crudest fashion: 
bubbles (bags) of perturbative vacuum in a confining medium, including eventual $O(\alpha_{s})$ corrections. Asymptotic freedom 
is implemented by considering free quarks and gluons inside color singlet bags, whereas confinement is realized by imposing 
that the vector current vanishes on the boundary.

Then, confinement is achieved by assuming a constant energy density for the vacuum (negative pressure), encoded in the 
so-called bag constant $B$, a phenomenological parameter extracted from fits to hadron masses. $B$ can also be viewed as 
the difference in energy density between the QCD and the perturbative vacua. A hadron energy (for a spherical bag) receives 
contributions from the vacuum and the kinetic energy, so that its minimum yields
\begin{equation}
E^{\rm min}_{h}=\frac{16}{3}\pi R_{h}^{3}B ~,
\end{equation}
and the hadron pressure (at equilibrium)
\begin{equation}
p_{h}=\frac{\partial E_{h}}{\partial V}=-B + \frac{{\rm const}}{4\pi R^{4}}=0 ~.
\end{equation}

Assuming the existence of a deconfining transition, the pressure in the quark-gluon plasma phase within this model is given by
\begin{equation}
p_{\rm QGP}=\left( \nu_{b}+\frac{7}{4}\nu_{f}\right) \frac{\pi^{2}T^{4}}{90}-B ~,
\end{equation}
whereas the pressure in the hadronic phase (taking, for simplicity, a pion gas) is given by
\begin{equation}
p_{\pi}=\nu_{\pi} \frac{\pi^{2}T^{4}}{90} ~,
\end{equation}
neglecting masses. Here, we have the following numbers of degrees of freedom:  $\nu_{\pi} = 3$, 
$\nu_{b} = 2 (N_{c}^{2} - 1)$ and $\nu_{f} = 2 N_{c}N_{f}$ for pions, gluons and quarks, respectively.

\begin{figure}[ht]
\begin{center}
\hspace*{-2.5cm}\includegraphics[width=8cm]{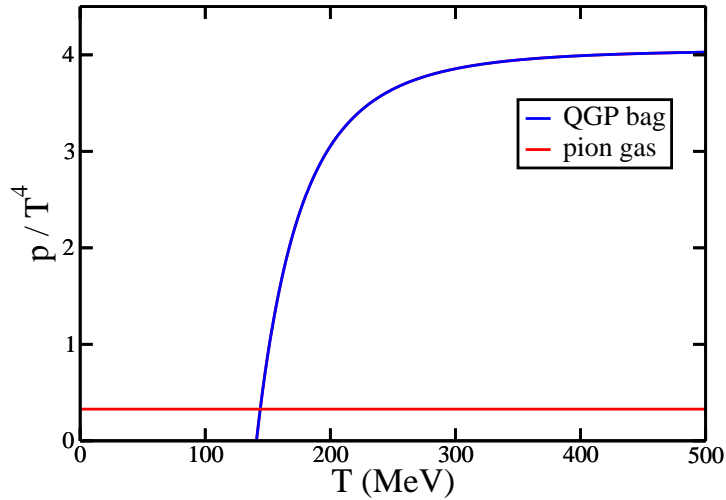}
\caption{Pressures in the bag model description.}
\label{fig:bag}
\end{center}
\end{figure}

For instance, for $N_{c} = 3$ , $N_{f} = 2$ and $B^{1/4} = 200~$MeV, we obtain the following critical temperature:
\begin{equation}
T_{c}= \left( \frac{45B}{17\pi^{2}} \right) \approx 144~ {\rm MeV}
\end{equation}
and a first-order phase transition as is clear from Fig. \ref{fig:bag}. The value of the critical temperature is actually 
very good as compared to recent lattice simulations \cite{lattice-Tc}, considering that this is a very crude model. On the 
other hand the nature of the transition, a crossover, is almost by construction missed in this approach.

\vspace*{-3.5cm}

\subsection{Basics of effective model building in QCD}

To go beyond in the study of the phases of QCD, one needs to know its symmetries, and how they are broken spontaneously 
or explicitly. But QCD is very involved. First, it is a non-abelian $SU(N_{c})$ gauge theory, with gluons living in the adjoint 
representation. Then, there are $N_{f}$ dynamical quarks who live in the fundamental representation. On top of that, these 
quarks have masses which are all different, which is very annoying from the point of view of symmetries.
So, in studying the phases of QCD, we should do it by parts, and consider many ``cousin theories'' which are very similar to 
QCD but simpler (more symmetric). In so doing, we can also study the dependence of physics on parameters which are fixed in Nature.

\begin{figure}[ht]
\begin{center}
\includegraphics[angle=90,width=10cm]{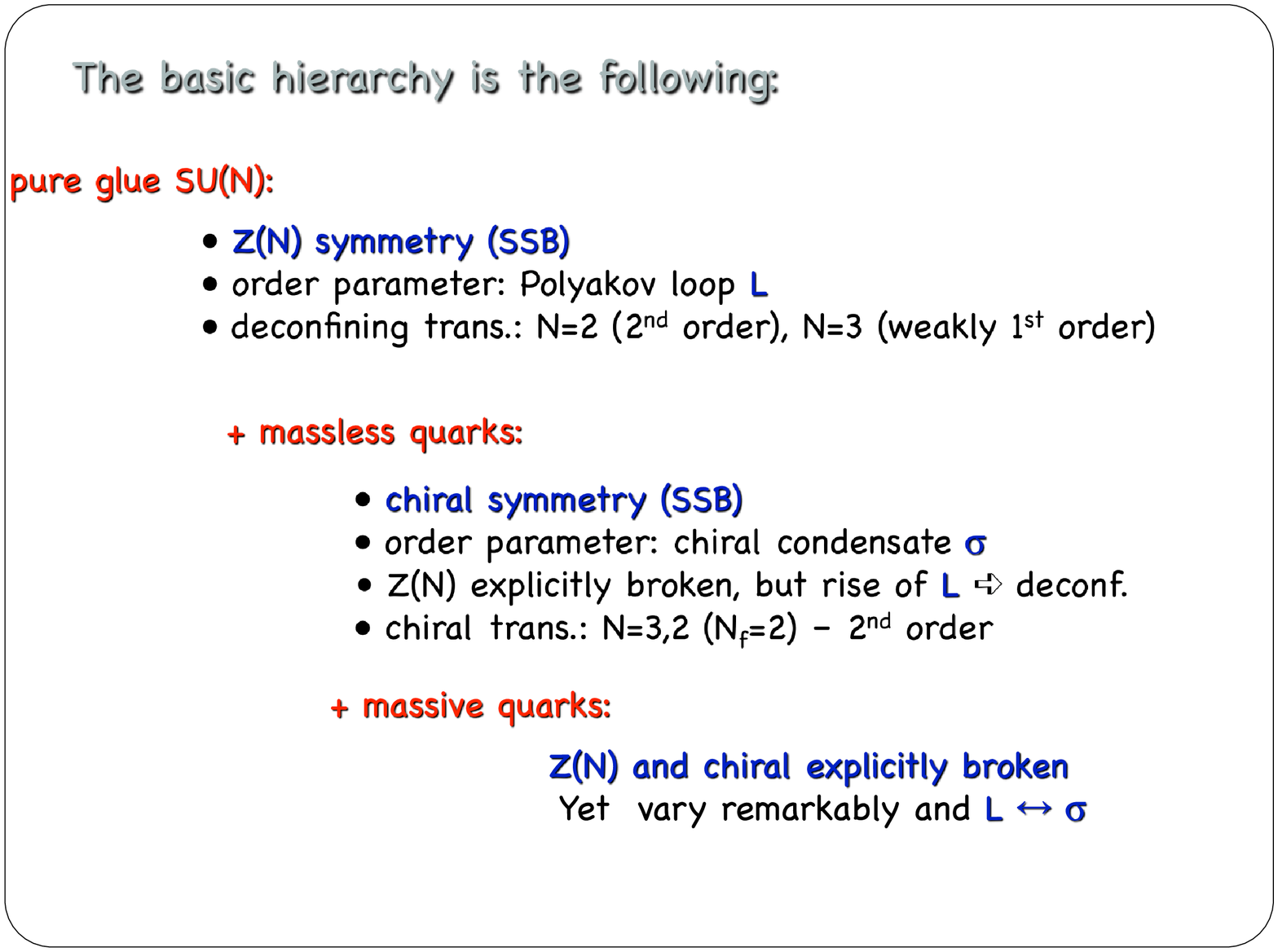}
\caption{Basic hierarchy in the step-by-step approach to QCD.}
\label{fig:hierarchy}
\end{center}
\end{figure}

Fig. \ref{fig:hierarchy} illustrates the step-by-step process one can follow in assembling the symmetry features present in QCD and 
learning from simpler theories, as well as cousin theories. Notice that the full theory, whose parameters are given by comparison to the 
experimental measurements, has essentially no symmetry left. Yet, some symmetries are mildly broken so that a ``memory'' of them 
remains. This fact allows us to use ``approximate order parameters'', for instance, a concept that is very useful in practice to characterize 
the chiral and deconfinement transitions.

\subsection{$SU(N_{c})$, $Z(N_{c})$ and the Polyakov loop}

In the QCD Lagrangian with massless quarks,
\begin{eqnarray}
{\cal L}&=&\frac{1}{2}{\rm Tr}F_{\mu\nu}F^{\mu\nu} + 
\bar q i\gamma^{\mu}D_{\mu}q  ~, \\
D_\mu &\equiv& (\partial_\mu - ig A_\mu)  ~, \\
F_{\mu\nu}&=&\frac{i}{g}~[D_\mu(A),D_\nu(A)] ~,
\end{eqnarray}
we have invariance under local $SU(N_{c})$. In particular, we have invariance under elements of the center group $Z(N_{c})$ 
(for a review, see Ref. \cite{rob-cargese})
\begin{eqnarray}
\Omega_c = e^{i\frac{2n\pi}{N_c}}{\bf 1} ~.
\end{eqnarray}
At finite temperature, one has also to impose the following boundary conditions:
\begin{eqnarray}
A_\mu(\vec{x},\beta)&=&+A_\mu(\vec{x},0) ~,\\
q(\vec{x},\beta)&=&-q(\vec{x},0) ~.
\end{eqnarray}
Any gauge transformation that is periodic in $\tau$ will do it. However, `t Hooft noticed that the class of possible transformations 
is more general. They are such that
\begin{eqnarray}
\Omega(\vec{x},\beta)=\Omega_{c} \quad , \quad \Omega(\vec{x},0)={\bf 1} ~,
\end{eqnarray}
keeping the gauge fields invariant but not the quarks.

For pure glue this $Z(N_{c})$ symmetry is exact and we can define an order parameter - the Polyakov loop:
\begin{eqnarray}
L(\vec{x})=\frac{1}{N_{c}}{\rm Tr} ~ {\cal P}\exp \left[ ig\int_0^\beta d\tau ~\tau^a A_0^a (\vec{x},\tau) \right] ~,
\end{eqnarray}
with $L$ transforming as
\begin{eqnarray}
L(\vec{x})\mapsto \Omega_c ~L(\vec{x})~ {\bf 1} = e^{i\frac{2n\pi}{N_c}}L(\vec{x})~.
\end{eqnarray}
At very high temperatures, $g \sim 0$, and $\beta \mapsto 0$, so that
\begin{eqnarray}
\langle\ell\rangle = e^{i\frac{2n\pi}{N_c}} \ell_{0} \quad , \quad  \ell_{0} \sim 1 ~,
\end{eqnarray}
and we have a $N$-fold degenerate vacuum, signaling spontaneous symmetry breaking of global $Z(N_{c}$). 
At $T = 0$, confinement implies that $\ell_{0} = 0$. Then, $\ell_{0} = 0$ can be used as an order parameter for the 
deconfining transition:
\begin{eqnarray}
\ell_{0}=0 ~,~ T<T_{c} \quad ; \quad \ell_{0}>0 ~,~ T>T_{c} ~.
\end{eqnarray}
Usually the Polyakov loop is related to the free energy of an infinitely heavy test quark via (confinement: no free quark)
\begin{eqnarray}
\langle\ell\rangle = e^{-F_{test}/T} ~.
\end{eqnarray}
See, however, the critical discussion in Ref. \cite{rob-cargese}.

\begin{figure}[ht]
\begin{center}
\includegraphics[width=8cm]{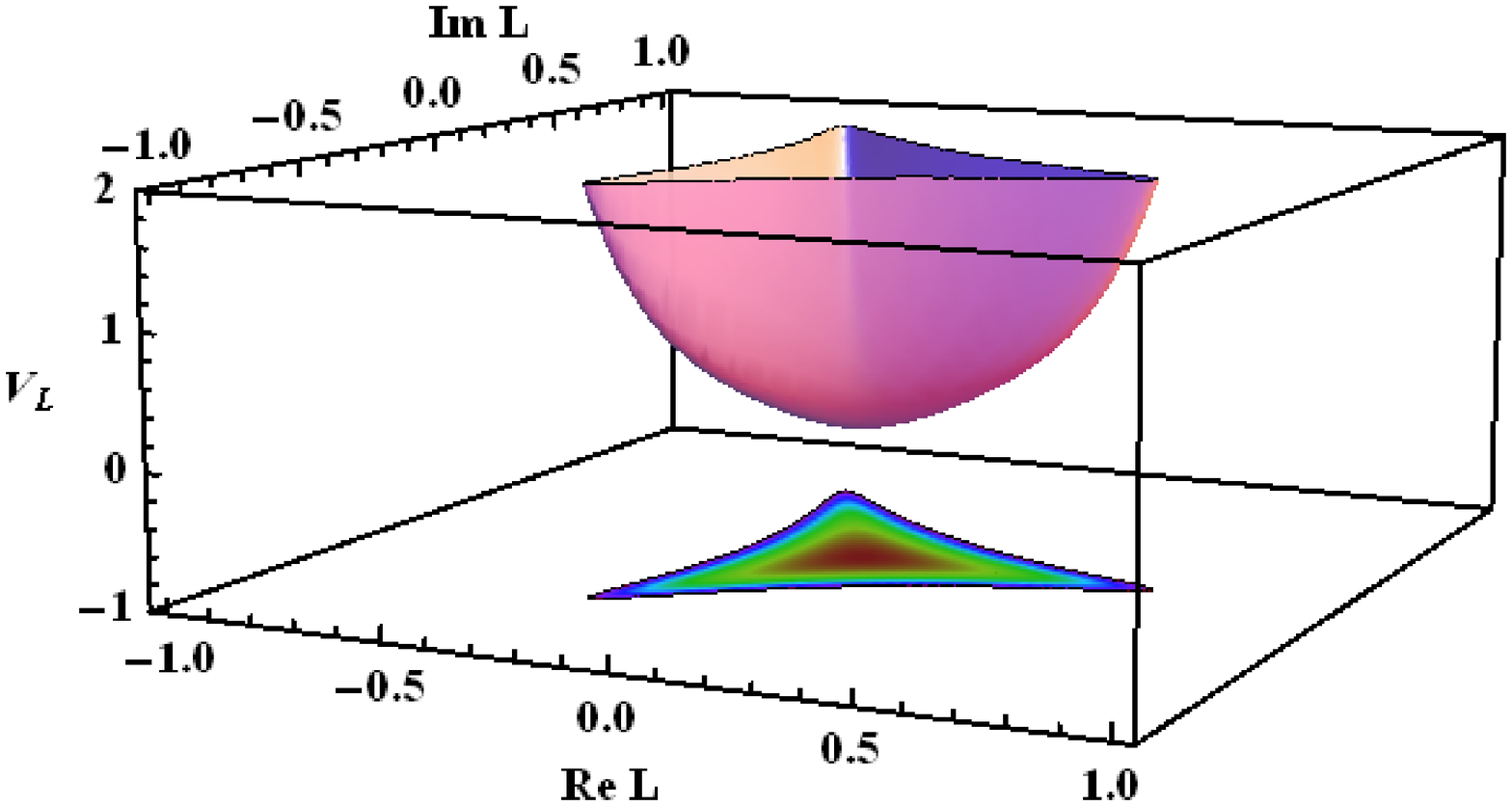}
\includegraphics[width=8cm]{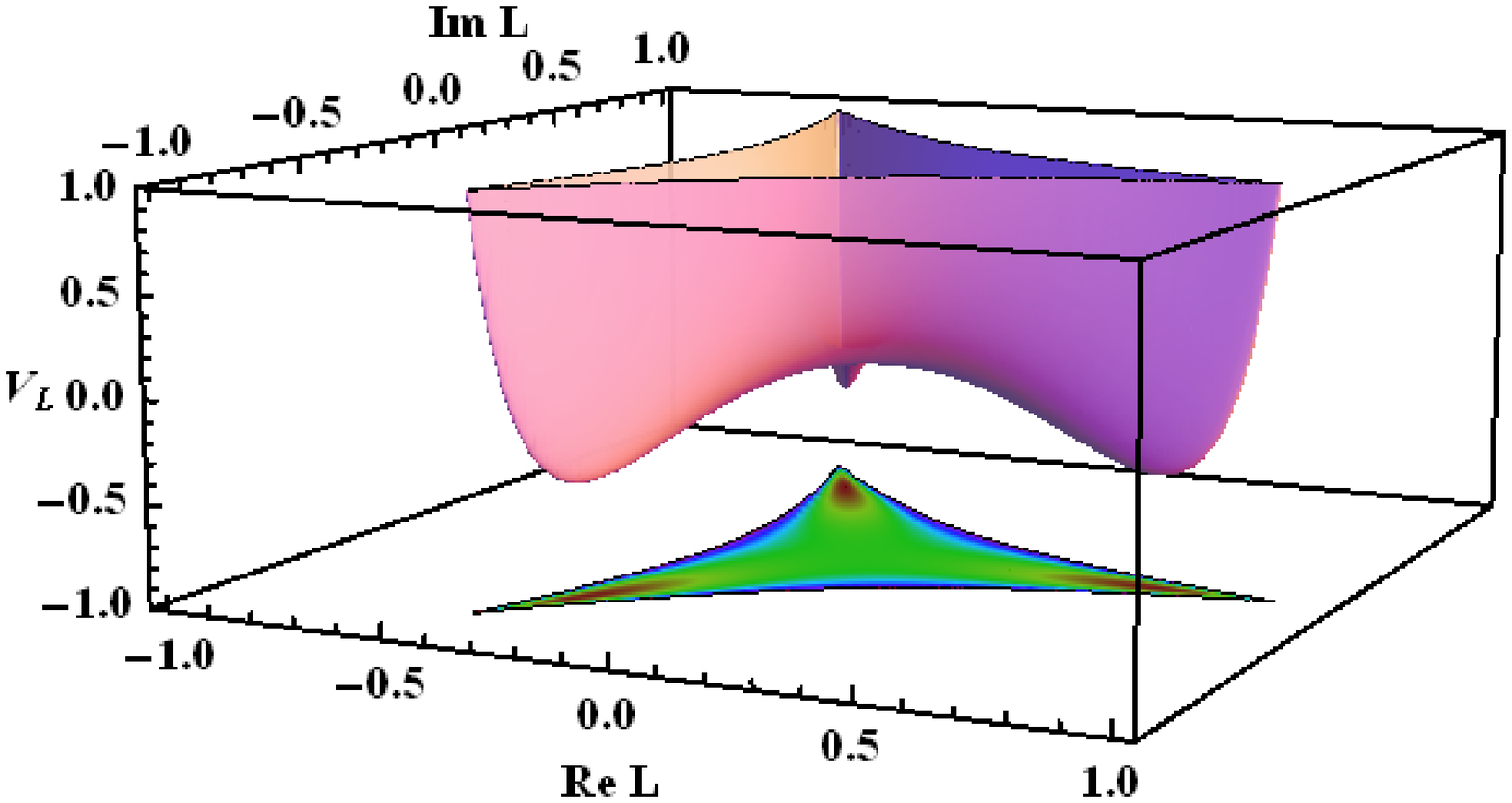}
\caption{Effective potential for the Polyakov loop for $T<T_{c}$ (upper) and $T>T_{c}$ (lower). Extracted from Ref. \cite{Mizher:2010zb}.}
\label{fig:pot-poly}
\end{center}
\end{figure}

The analysis above is valid only for pure glue, i.e. with no dynamical quarks. However, we can still ask whether $Z(3)$ is an 
approximate symmetry in QCD. On the lattice, in full QCD, one sees a remarkable variation of $\ell$ around $T_{c}$, so that it 
plays the role of an approximate order parameter \cite{lattice-polyakov-chiral}. 
Notice, however, that $Z(3)$ is broken at high, not low $T$, just the opposite 
of what is found in the analogous description of spin systems, such as Ising, Potts, etc \cite{BCS-BEC}. 
The effective potential for the Polyakov loop is illustrated in Fig. \ref{fig:pot-poly}.

\subsection{Adding quarks: chiral symmetry}

In the limit of massless quarks, QCD is invariant under global chiral rotations $U(N_{f})_{L} \times U(N_{f})_{R}$ 
of the quark fields. One can rewrite this symmetry in terms of vector ($V = R + L$) and axial ($A = R - L$) rotations 
\begin{eqnarray}
U(N_{f})_{L} \times U(N_{f})_{R} \sim U(N_{f})_{V}\times U(N_{f})_{A} ~.
\end{eqnarray}
As $U(N) \sim SU(N) \times U(1)$, one finds
\begin{eqnarray}
U(N_{f})_{L} \times U(N_{f})_{R} \sim SU(N_{f})_{L} \times SU(N_{f})_{R} \times U(1)_{V} \times U(1)_{A} ~,
\end{eqnarray}
where we see the $U(1)_{V}$ from quark number conservation and the $U(1)_{A}$ broken by instantons.

In QCD, the remaining $SU(N_{f})_{L} \times SU(N_{f})_{R}$ is explicitly broken by a nonzero mass term. 
Take, for simplicity, $N_{f}=2$. Then,
\begin{eqnarray}
{\cal L}=\frac{1}{4}F^{a}_{\mu\nu}F^{a\mu\nu} + 
\overline{\psi}_{L} \gamma^{\mu}D_{\mu} \psi_{L} +
\overline{\psi}_{R} \gamma^{\mu}D_{\mu} \psi_{R}-
m_{u}(\overline{u}_{L} u_{R} + \overline{u}_{R} u_{L})-
m_{d}(\overline{d}_{L} d_{R} + \overline{d}_{R} d_{L} ) ~,
\end{eqnarray}
so that, for non-vanishing $m_{u}= m_{d}$, the only symmetry that remains is the vector isospin $SU(2)_{V}$. 
In the light quark sector of QCD, chiral symmetry is just approximate. Then, for massless QCD, one should find parity 
doublets in the vacuum, which is not confirmed in the hadronic spectrum. Thus, chiral symmetry must be broken in the 
vacuum by the presence of a quark chiral condensate, so that 
\begin{eqnarray}
SU(N_{f})_{L} \times SU(N_{f})_{R} \mapsto SU(N_{f})_{V} ~,
\end{eqnarray}
and the broken generators allow for the existence of pions, kaons, etc.

Hence, for massless QCD, we can define an order parameter for the spontaneous breaking of chiral symmetry in the vacuum 
- the chiral condensate:
\begin{eqnarray}
\langle 0 | \overline{\psi}\psi | 0 \rangle=
\langle 0 | \overline{\psi}_{L}\psi_{R} | 0 \rangle + \langle 0 | \overline{\psi}_{R}\psi_{L} |0 \rangle ~,
\end{eqnarray}
so that this vacuum expectation value couples together the $L$ and $R$ sectors, unless in the case it vanishes. For very high 
temperatures or densities (low $\alpha_{s}$), one expects to restore chiral symmetry, melting the condensate that is a function 
of $T$ and quark masses and plays the role of an order parameter for the chiral transition in QCD.

Again, the analysis above is valid only for massless quarks. However, we can still ask whether QCD is approximately chiral 
in the light quark sector. On the lattice (full massive QCD), one sees a remarkable variation of the chiral condensate around $T_{c}$, 
so that the condensate plays the role of an approximate order parameter \cite{lattice-polyakov-chiral}.

In summary, there are two relevant phase transitions in QCD, associated with spontaneous symmetry breaking mechanisms 
for different symmetries of the action: (i) an approximate $Z(N_{c})$ symmetry and deconfinement, which is exact for pure gauge 
$SU(N_{c})$ with an order parameter given by the Polyakov loop; (ii) an approximate chiral symmetry and chiral transition, which 
is exact for massless quarks, with an order parameter given by the chiral condensate. 

One can try to investigate these phase transitions by building effective models based on such symmetries of the QCD action. 
Then, the basic rules would be: 
(i) keeping all relevant symmetries of the action; (ii) trying to include in the effective action all terms allowed by the chosen symmetries; 
(iii) developing a mimic of QCD at low energy using a simpler field theory; (iv) providing, whenever possible, analytic results at least 
for estimates and qualitative behavior. Well-known examples are the linear sigma model, the Nambu-Jona-Lasinio model, 
Polyakov loop models and so on \cite{QGP}. Although they represent just part of the story, combined with lattice 
QCD they may provide good insight.

\section{A final comment}

Instead of conclusions, just a final comment on a point we have already made in the discussion above. 
To make progress in understanding, or at least in collecting facts about, (de)confinement and chiral symmetry, we need it all: 
experiments and observations, lattice simulations, theory developments, effective models, and also combinations whenever possible. 
In that vein, it is absolutely crucial to have theorists and experimentalists working and discussing together. 

\section*{Acknowledgements}

The work of ESF was financially supported by the Helmholtz International Center for FAIR 
within the framework of the LOEWE program (Landesoffensive zur Entwicklung Wissenschaftlich-\"Okonomischer 
Exzellenz) launched by the State of Hesse.



\end{document}